\documentclass[10pt,letterpaper]{article}
\usepackage[top=0.85in,left=2.75in,footskip=0.75in]{geometry}

\usepackage{amsmath,amssymb}

\usepackage{changepage}

\usepackage[utf8x]{inputenc}

\usepackage{textcomp,marvosym}

\usepackage{cite}

\usepackage{nameref,hyperref}

\usepackage{microtype}
\DisableLigatures[f]{encoding = *, family = * }

\usepackage[table]{xcolor}

\usepackage{array}

\newcolumntype{+}{!{\vrule width 2pt}}

\newlength\savedwidth

\raggedright
\usepackage[aboveskip=1pt,labelfont=bf,labelsep=period,justification=raggedright,singlelinecheck=off]{caption}

\makeatletter
\renewcommand{\@biblabel}[1]{\quad#1.}
\makeatother

\usepackage{lastpage,fancyhdr,graphicx}
\usepackage{epstopdf}
\fancyheadoffset[L]{2.25in}
\fancyfootoffset[L]{2.25in}
\begin{document}
\vspace*{0.2in}

\begin{flushleft}
{\Large
\textbf\newline{Linking Labs: Interconnecting Experimental Environments} 
}
\newline
\\
Tanja Schultz\textsuperscript{1*},
Felix Putze\textsuperscript{1},
Thorsten Fehr\textsuperscript{2},
Moritz Meier\textsuperscript{1},
Celeste Mason\textsuperscript{1},
Florian Ahrens\textsuperscript{2},
Manfred Herrmann\textsuperscript{2}
\\
\bigskip
\textbf{1} Cognitive Systems Lab, University of Bremen, Germany
\\
\textbf{2} Department of Neuropsychology and Behavioral Neurobiology, University of Bremen, Germany
\\
\bigskip

%
%





* tanja.schultz@uni-bremen.de

\end{flushleft}
\section*{Abstract}
We introduce the concept of \textit{LabLinking}: a technology-based interconnection of experimental laboratories across institutions, disciplines, cultures, languages, and time zones - in other words \textit{experiments without borders}. In particular, we introduce LabLinking levels (LLL), which define the degree of tightness of empirical interconnection between labs. We describe the technological infrastructure in terms of hard- and software required for the respective LLLs and present examples of linked laboratories along with insights about the challenges and benefits. In sum, we argue that linked labs provide a unique platform for a continuous exchange between scientists and experimenters, thereby enabling a time synchronous execution of experiments performed with and by decentralized user and researchers, improving outreach and ease of subject recruitment, allowing to establish new experimental designs and to incorporate a panoply of complementary biosensors, devices, hard- and software solutions.      



\section{Introduction}
Human behavior and human-decision making are frequently studied in laboratory experiments. Typical experiments leverage continuous recordings of participants' speech, motion, brain activity, and other processes, resulting in a variety of high-dimensional biosignals \cite{schultz2017biosignal}. Due to the range and complexity of signals, setups, devices, and research questions, laboratories often specialize in the study of certain aspects of human behavior or capturing of particular biosignals~\cite{haidu_automated_2019, meier2018}, or behavioral responses~ \cite{marsh_is_2013,fehr_neural_2018,peukert_acceptance_2019}. While specialization is often the prerequisite for mastering a task, it can also reduce options in the sense that researchers are bound to the available hardware and software, and thus are limited to certain modalities and biosignals, to investigation and analysis approaches, to physical environments, to the amount of concurrently recordable participants, and to the skills, features, and language of participants, to name only a few potential limitations.

We want to overcome these limitations by introducing a new experimental approach termed \emph{LabLinking}.
LabLinking refers to interconnecting people in spatially distributed labs, who participate in joint experiments while sharing a common task.

Without loss of generality, we introduce examples of LabLinking to study various interaction forms and decision-making situations. 
By tightly interconnecting labs with each other, more complementary sensors and measuring devices become available, synergies of local expertise can be leveraged, and larger study samples could be achieved by attracting participants to the distributed locations. In addition, the distributed locations of the linked labs allow for qualitatively new perspectives, since the impact of different time zones, languages, environments etc.\ to the forms of interaction also might be approached in an ecologically valid sense.

\section{Linking Labs and People}
Labs can be linked on different levels of tightness, we refer to this as \textit{Lab-Linking-Level (LLL)}. With increasing LLL the linking quality improves with respect to richness and quality of participants' interaction, task sharing, complexity of decisions to be taken, as well as time synchronicity. Consequently, the requirements on the linking technology will grow - as will the possibilities for synergies and benefits for the members of linked labs. In the following sections, we first define the range of LLLs and the roles, which participants of lab experiments may take, and then show initial experiments and results achieved in practice via lab linking. 

\subsection{LabLinking-Level (LLL)}
\noindent \textbf{LLL-1: Coordinated studies}. Labs agree to collaborate by exchanging experience, best practices, and lessons learnt, by discussing protocols, and by using compatible equipment. They establish common terminology, data formats, annotation schemes, and documentations for the purpose of sharing (common or complementary) data and experimental paradigms. Typically, LLL-1 connected labs carry out joint projects with same scenario but separate experiments. \emph{Requirements}: Willingness of partners to adjust their procedures to common protocols. \emph{Benefits}: Leveraging complementary knowledge and insights, learning from each other and exchanging experience often results in synergies, which make the whole more than the sum of its parts. \\

\noindent \textbf{LLL-2: Asynchronous data coupling}. Beyond coordination, researchers from LLL-2 connected labs study the same experiment from different perspectives and equipment (e.g., different modalities, sensors). While data are recorded separately, for example when same stimuli are presented to participants of two participating labs, the data will be synchronized afterwards through means of data synchronization. Consequently, data are treated as asynchronous but parallel recordings and analyzed accordingly. An example of LLL-2 collaboration takes place in the research area \textit{Human Modeling} within the CRC 1320 Everyday Activity Science and Engineering (EASE), as described in Section~\ref{sec:example}. \emph{Requirements}: Hard- and software to perform synchronization of data streams, using for example LabStreamingLayer (\url{https://github.com/sccn/labstreaminglayer}) or hardware synchronization via light sensors or serial buses. \emph{Benefit}: Details and insights from complementary data, modalities, and spatially distributed perspectives which give two sides of the same coin.\\

\noindent \textbf{LLL-3: Synchronous bi-directional data coupling}. Level LLL-3 allows moving from asynchronous to synchronous experiment designs, which combine one or more recordings within the same spatially distributed setup. A low-latency "real-time" synchronization via internet connection allows interaction between the involved participants and labs. \emph{Requirements}: Stable, high-throughput network connection, ability for synchronized and coordinated execution of experiments. \emph{Benefit}: Allows performing interaction scenarios between labs (e.g., involving decision making or consensus finding). 
\\

\noindent \textbf{LLL-4: Immersive synchronous interaction and collaboration}. Level LLL-4 increases the degree of immersion and the interaction bandwidth between participants to create the impression of a co-located experiment, despite spatial separation. This is achieved through Virtual and Augmented Reality as well as multimodal tracking and interpretation of the participants' behavior and state. \emph{Requirements}: Complex experiment setup for tracking and measuring participants as well as for creating the virtual laboratory environment. \emph{Benefit}: Spatially distributed real-time experiments feel similar to co-located experiments and yield valid results.
\\

\noindent \textbf{LLL-5+: LabLinking  based on future emerging technologies}. LLLs are expected to increase over time as the tightness level of LabLinking will likely benefit from future emerging technologies, for example through distributed physical interaction which feels almost real. Thus, spatial distance will become less noticeable and boundaries between labs will start to blur. Such future developments will pave the way for large-scale LabLinking between numerous labs across the world.\\

\subsection{Participants' Roles}
Independent of the above described LabLinking level, participants can have different roles within an experiment. To characterize the different roles, we formulate three aspects: 1) \emph{agency} considers if the participant has the ability to make their own decisions and have them influence the course of the experiment. 2) \emph{linkage} considers if the participant has the ability to influence/be influenced by another participant or whether they are acting independently. 3) \emph{placement} considers whether the participant is situated inside the experiment scene or takes part from an outside perspective.

\begin{figure}[ht]
\centering
\includegraphics[width=\textwidth]{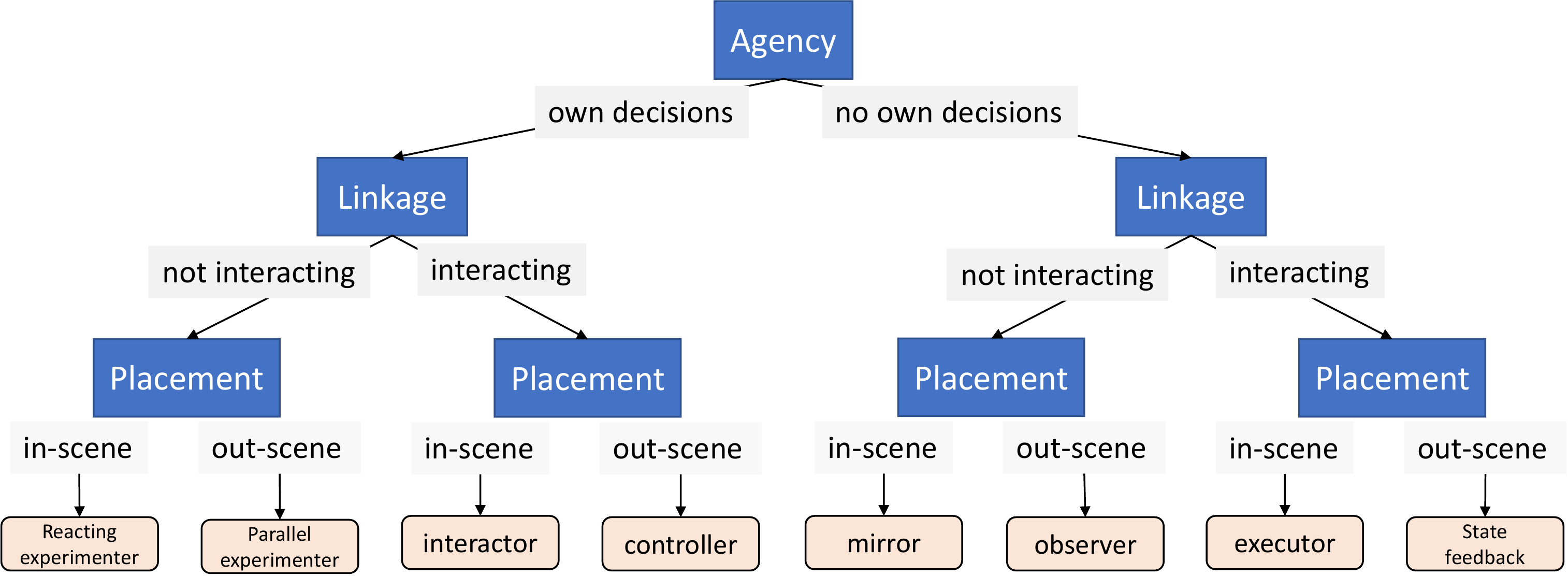}
\caption{Overview of possible participant roles in different LabLinking configurations.}
\label{fig:roles}
\end{figure}

As illustrated in Figure~\ref{fig:roles}, the different combination of role characteristics can be combined to certain role prototypes. The depicted tree goes from relation of the participant to themselves (agency), to the relation to other participants (linkage), and finally to their relation to the environment.
We will go through all role prototypes with examples in the following:

\begin{enumerate}
    \item \textbf{Reacting experimenter:} Such a participant is observing another participant. S/he acts upon his/her own scene which is not influenced by the observed participant. For example, a kid (reacting experimenter) learns to set the dinner table by watching and imitating an adult in a remote setting.
    \item \textbf{Parallel experimenter:} Here, participants act independently of each other without the means of coordinating with each other. This is the only role available in LLL1. For example, two participants perform the same task in different labs and data is annotated according to a joint annotation scheme.
    \item \textbf{Interactor:} In this role, the participant influences the experiment for others directly or indirectly. The interaction itself does not need to be symmetrical, i.e., the selection of actions, the times at which an action is possible, etc.\ can vary between participants (requires at least LLL-3). For example, the interactor sends recommendations to another participant through an microphone and earpiece.   
    \item \textbf{Controller:} Such participant can be considered to be in a ``control room'' influencing the scene, e.g., through triggered events, without being part of it and without being restricted to the perception capabilities of an agent in the scene. For example, the controller sends buttons to highlight task actions which another participant is instructed to execute.
    \item \textbf{Mirror:} Here, the passive participant is asked to imagine being in the shoes of the active participant. This is the role configuration which was employed in the offline EASE data collection, see Section~\ref{sec:lll2_example} (requires at least LLL-2). For example, a mirror participant is watching a video feed from the point-of-view camera of another participant and imagines performing the same task.
    \item \textbf{Observer:} In this configuration, the participant is still passive, but is now asked to observe, judge, remember, or otherwise process the actions of the active participant in the other laboratory (requires at least LLL-2). For example, an observer watches another participant performing a task and is judging their actions.
    \item \textbf{Executor:} A participant in this role executes actions within the scene, but has no agency over them, e.g., because they are determined and communicated by another participant. For example, the other participant in the "Controller" example has the role of Executor.
    \item \textbf{State Feedback:} A participant in this role influences the experiment scene they observe unconsciously, as it reacts to their measured (cognitive or affective) state. For example, a participant's engagement level is measured while they observe another participant executing a task. If engagement falls below a threshold, the task speed is increased.
\end{enumerate}

\subsection{LabLinking Ecosystems}
Linking labs and people is a very general concept and neither limited to the number of interconnected labs and participants, nor is it bound to physical time or space. 

First, LabLinking allows experiments to be performed concurrently, bringing together different configurations of decentralized users, i.e. individuals, dyads, triads, or larger groups and teams of participants. 
Second, LabLinking can take place in reality and/or in a common virtual or augmented reality AR/VR). Consequently, labLinked decision-making situations could be detected and investigated in a multi-dimensional manner by complementary measuring instruments available in various laboratories, the synergies of local expertise could be combined, and larger study volumes with participants from different locations could be achieved. Also, embracing AR/VR technology ensures that LabLinking will benefit from the expected disruptive advancements of this technology in the years to come. Third, the remote physical distance between the participating labs allows qualitatively new perspectives, since differences between the forms of interaction can be ecologically validated. 
Last but not least, LabLinking is a very suitable and applicable concept for the teaching and education of students and early career scientists across disciplines. Teaching could happen without borders, and remotely in times when people cannot meet face-to-face. Thus, laboratory and software should be routinely integrated into massive online education and training in the form of exercises, internships, theses, and project studies.

\section{LabLinking in Practice}\label{sec:example}
In this section, we present practical examples of LabLinking that allow to observe people performing everyday activities in a novel way that would not have been possible without the concept of LabLinking. In particular, we describe the benefits of the simultaneous recordings of biosignal data corresponding to everyday activities. These recordings leverage several modalities captured by various sensors, ranging from wearable low-cost acceleration sensors to high-dimensional costly functional brain imaging devices \cite{mason_iros2020}. 

\subsection{Participating Labs}
In this first phase of LabLinking, we have successfully linked two participating labs already (and are getting ready to link in another two labs soon), where one lab focuses on AR/VR and another experienced in synchronized acquisition of data from large subject groups. We begin this section by introducing the two labs at the core of our long-term LabLinking strategy. 

\subsubsection{Biosignals Lab @ CSL} 
\begin{figure}[ht]
\centering
\includegraphics[width=0.6\columnwidth]{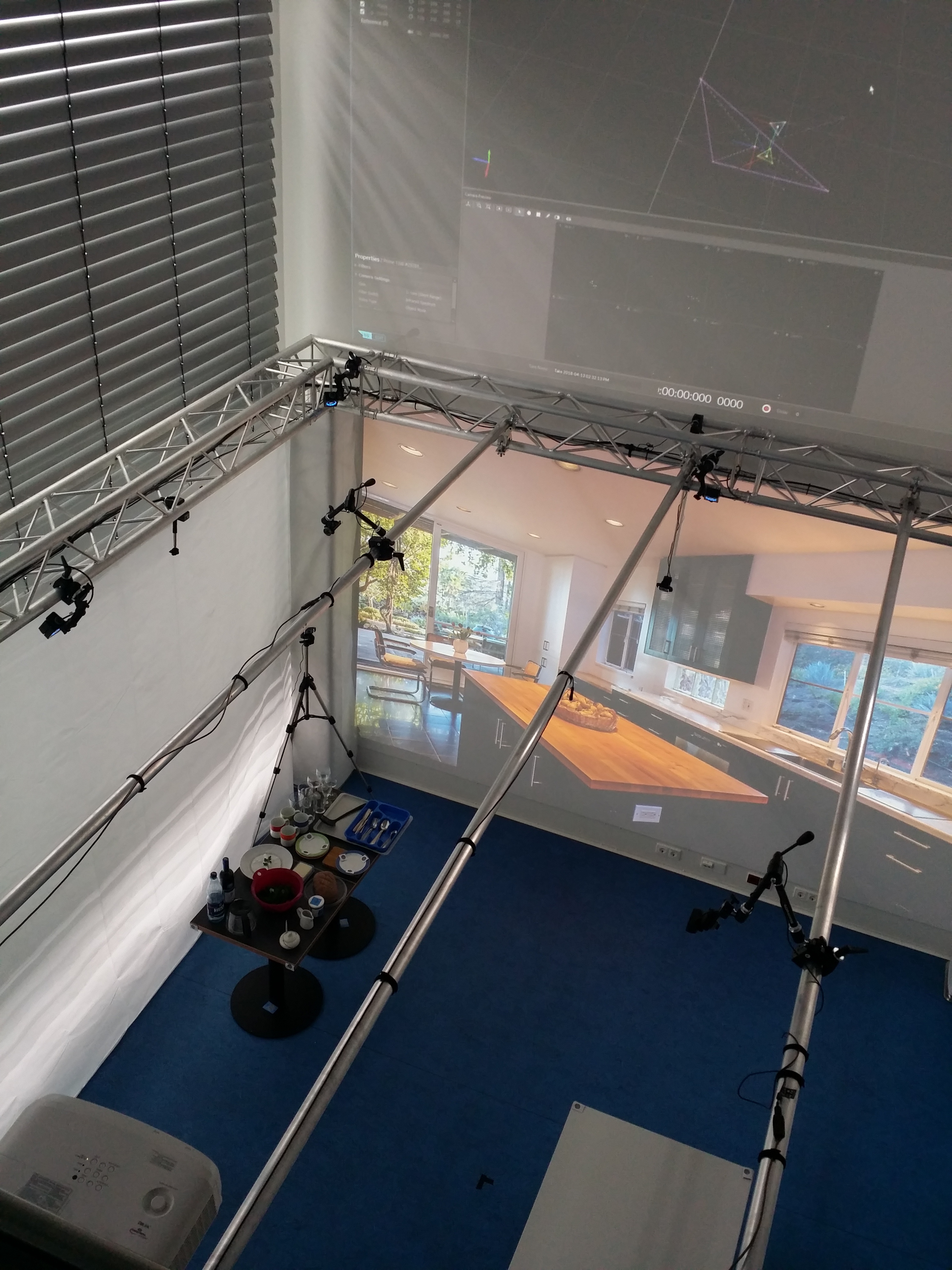}
\caption{The Biosignals Lab at the Cognitive Systems Lab (CSL), University Bremen}
\label{fig:BASE}
\end{figure}

The Biosignals Lab at the Cognitive Systems Lab (CSL) consists of an interaction space (5x4m) which allows to blend real with virtual reality interactions (see Figure \ref{fig:BASE}). The Biosignals lab is fully equipped with a range of sensory devices to capture biosignals resulting from human behaviour like speech, motion, eye gaze, muscle and brain activities under both controlled and less restricted open-space conditions. The sensors, devices, and equipment available include Hololenses, stationary and head-mounted cameras, near- and far-field microphones for speech and audio event recording, a marker-based 9-camera OptiTrack motion capture system, wireless motion tracking based on PLUX (\url{https://plux.info/}) inertial sensors, electrodermal activity (EDA) sensors, mobile eye-tracking with Pupil Labs (\url{https://pupil-labs.com/}) headsets, muscle activity acquisition with stationary 256-channel and mobile 4-channel electromyography (EMG) devices from OT Bioelettronica (\url{https://www.otbioelettronica.it}) and PLUX, and brain activity recording based on a BrainProducts (\url{https://www.brainproducts.com/}) actiCHamp 64-channel electroencephalography (EEG) and mobile EEGs based on OpenBCI (\url{https://openbci.com/}) and g.Tec's g.Nautilus (\url{https://www.gtec.at}). See \cite{meier2018} for more details on the Biosignals Lab, the hard- and software setup as well as the various devices. Furthermore, the Biosignals Lab comprises a large shielding cabin to record high-quality biosignals in clean and controlled conditions. 

\subsubsection{The NeuroImaging and EEG-Lab}

The NeuroImaging and EEG-Lab is hosted by the Department of Neuropsychology and Behavioral Neurobiology and forms part of the Center of Advanced Imaging project (CAI). The CAI was equipped with a 3 Tesla Siemens Allegra® headscanner, followed by a 3 Tesla Siemens Skyra® scanner in a joined initiative with Fraunhofer MEVIS. The research focus on the neural correlates of executive control and different sort of conflict processing and interference resolution in complex decision making both in laboratory and semi-natural context~\cite{gloy2020decision}. Most investigations are performed using both fMRI and EEG devices~\cite{Trautmann-Lengsfeld2013}.

\begin{figure}[ht]
\centering
\includegraphics[width=0.8\columnwidth]{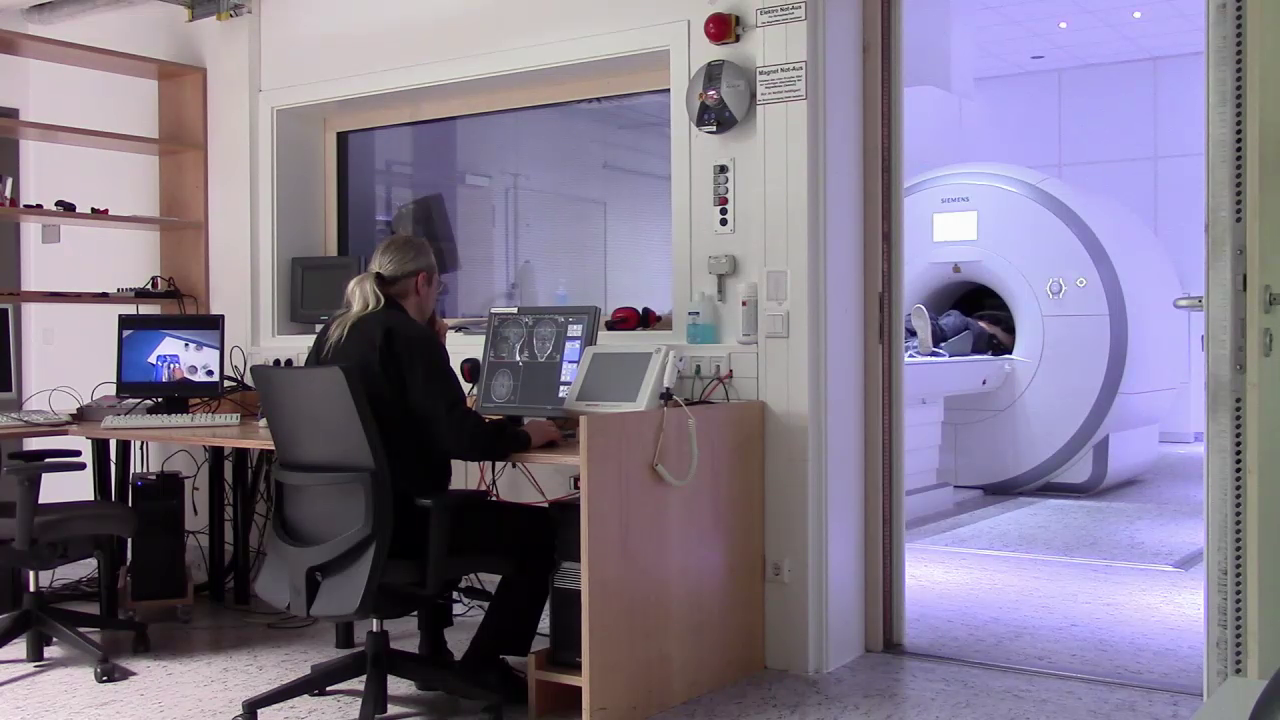}
\caption{The Neuroimaging Lab, fMRI measurement during video observation}
\label{fig:flo_MRT}
\end{figure}

\subsection{A Case of LLL-1 - Human Everyday Activities}
\begin{figure}[ht]
\centering
\includegraphics[width=0.8\columnwidth]{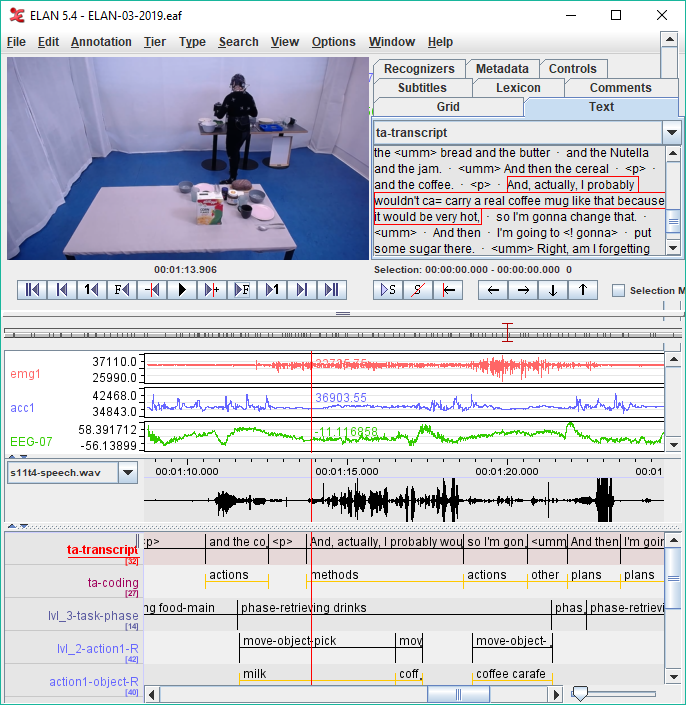}
\caption{LLL-1: Partitur file with standardized annotation and time-alignment of multi-modal biosignals.}
\label{fig:Partitur}
\end{figure}

Our first example describes LabLinking Level 1 (LLL-1), which is applied to the first phase of the 
collaborative research center 1320 ``Everyday Science and Engineering (EASE)'' http://ease-crc.org). EASE focuses on facilitation of robotic mastery of everyday activities as its unifying mission. For this purpose, we observe humans' performance of everyday activities using a multitude of sensors and devices. The resulting biosignals are derived from corresponding brain, muscle, and speech activities. Their analyses provide insights into complementary aspects of behavior required for humans to masterfully perform everyday activities with little effort or attention. Several labs, including the Biosignals and the NeuroImaging Labs, jointly investigate a common scenario, the table setting task. The participating labs study a variety of aspects through a panoply of modalities. A common annotation standard and an ontology form the means to integrate different levels of abstraction and diverse experiments. Selected manual annotation of data is applied to bootstrap semi-automatic annotation procedures based on the human-in-the-loop concept. Joint toolkits like EaseLAN (based on the ELAN Framework, \cite{wittenburg2006}) were created to support data visualization, annotation, and time-aligned arrangement in score files coined "Partitur", as shown in Figure~\ref{fig:Partitur}. By integrating results from a wide range of data sources and complex analyses using a multitude of complementary methods, we envision to effectively transfer an extensive contextually dense reserve of human everyday activity experience and problem solving approaches to robotic agents. A collection of high-dimensional biosignals data from about 100 participants along with rich and time-aligned annotations will be made available open-source to support common standards.

\subsection{A Case of LLL-2 - Everyday Activity in MRT}\label{sec:lll2_example}

Brain activity measurements using methods such as fMRI or EEG allow inspection of what people pay attention to as they perform tasks, how they may adapt to ambiguous situations or unforeseen obstacles, what might influence their decision making processes, and how their own motor imagery when viewing performance of activities compares with in-situ motor execution. In the Biosignals Lab, we already capture a 16-channel EEG with a mobile device attached to the head of the person executing the activity. This gives us important information, e.g., to infer the person's workload or to analyze motor activation during physical activities. However, physical activity impacts the EEG signal, also the spatial resolution of EEG is limited and the number of available channels does not allow robust source localization of brain activity during task execution. A higher density of electrodes would extend setup time beyond practicable limits. 

On LLL-2, prerecorded performances of everyday activities can therefore be used in neuroimaging studies at the NeuroImaging Lab to achieve a more comprehensive picture of brain activity. For this purpose, videos from the perspective of the acting person are recorded in Biosignals Lab via head mounted camera by an experimenter who acts out scenarios that resemble those of participants of table setting experiments. Scenarios encompass a variety of confidently finished runs as well as runs causing erroneous behavior induced by missing or misplaced objects, or plans prevented from execution. Visible actions consist of articulate and easily traceable movements of the arms and hands as well as smooth camera pans of the head to establish and validate a standardized set of videos. This standardized set is then presented to participants of EEG and fMRI studies at the NeuroImaging lab, who employ motor imagery~\cite{Jeannerod1995} to actively put themselves into the perceived scenes. Semantically unique episodes within the videos (e.g., pick, place) are manually annotated and the recorded brain activity of the participants is correlated with those episodes. 

EEG and fMRI are used as complementary tools in these viewing scenarios in order take advantage of the higher spatiotemporal signal information stemming from the integration of their individual data and to introduce a combined fMRI constrained source analysis~\cite{Trautmann-Lengsfeld2013}. Results reveal a wide range of activated brain networks with a high temporal and spatial resolution, which are further analyzed in terms of dimensionality of involved networks during the planning and execution of complex everyday activities and the handling of errors and ambiguous situations. Not only will this elucidate brain function during complex tasks for members of the neuroscience community, but it will also provide researchers in the field of robotics with a template of the organ whose function they are trying to mimic.

After successful validation of these standardized videos and further improvements of the automatic video annotation, later phases will likely allow participants to use their own datasets from previous active table setting experiments for an individually customized simulation of complex activities.

There are several pre-requisites to analyze the data in a synchronized fashion. These involve different aspects, such as: 1) Temporal synchronization, 3) Format synchronization, and 3) Semantic synchronization. 

ad 1) For temporal synchronization, i.e.\ the alignment of time series recorded from different sensors, modalities, and laboratories, we employ a network-based synchronization using the LabStreamingLayer architecture. LSL uses a clock synchronization protocol to locally align time stamps across different computers in a network. Compared to other synchronization options, such as synchronization through light sensors or serial port signals, network-based synchronization scales best when integrating devices with very different native protocols and technical platforms.

ad 2) For format synchronization, we need to consider common, documented ways of storing and transmitting the generated data. For this purpose, we rely on automated data processing pipeline which enforces desired data characteristics, such as video codec, sampling rate, channel ordering, etc. For storing data, we use the NEEM-Hub of EASE, which is a distributed storage for heterogeneous data. 

ad 3) For semantic synchronization, we need joint annotation schemes for manual and automatic annotation of the data streams (e.g., segments, classes, etc.). Semantic synchronization can be enforced by linking the annotations to an ontology, like SOMA (\url{https://ease-crc.github.io/soma/}), which precisely defines each concept used for annotation.

\begin{figure}[ht]
\centering
\includegraphics[width=\columnwidth]{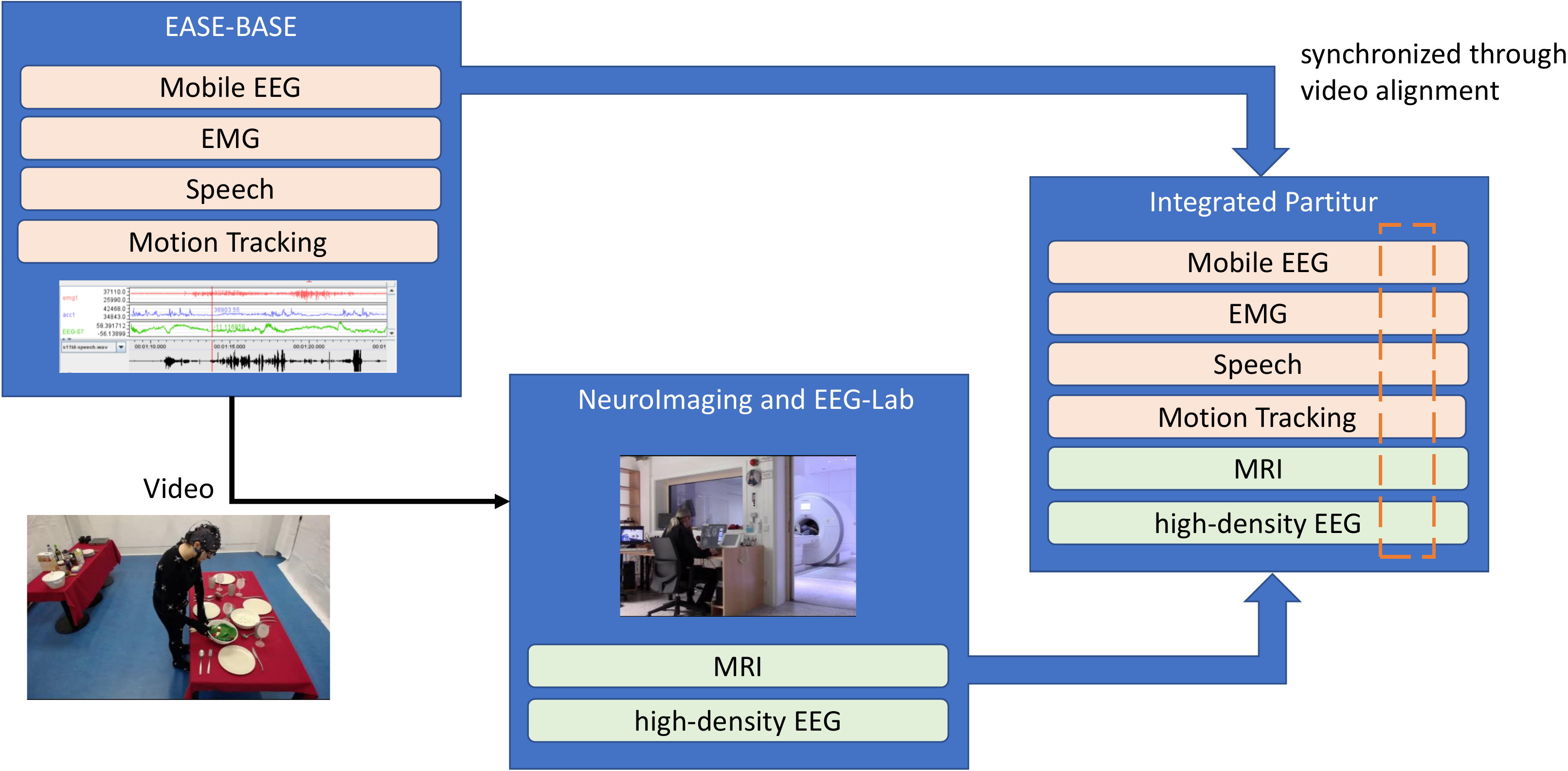}
\caption{LLL-2: Asynchronous / Offline LabLinking}
\label{fig:ll_offline}
\end{figure}

This analysis can be performed offline as described in Figure~\ref{fig:ll_offline}: First, multimodal data is recorded of a person setting the table, including videos from different perspectives. These videos are then shown to a person in the MRI scanner afterwards and the data is synchronized through the timestamps of the individual data samples in the video. However, this approach only works for LLL-2 but obviously is not sufficient once we want to expand the setup toward LLL-3+ for studying interactive behavior.

\subsection{A Case of LLL-3 - Real-Time LabLinking}

\begin{figure}[!ht]
\centering
\includegraphics[width=\columnwidth]{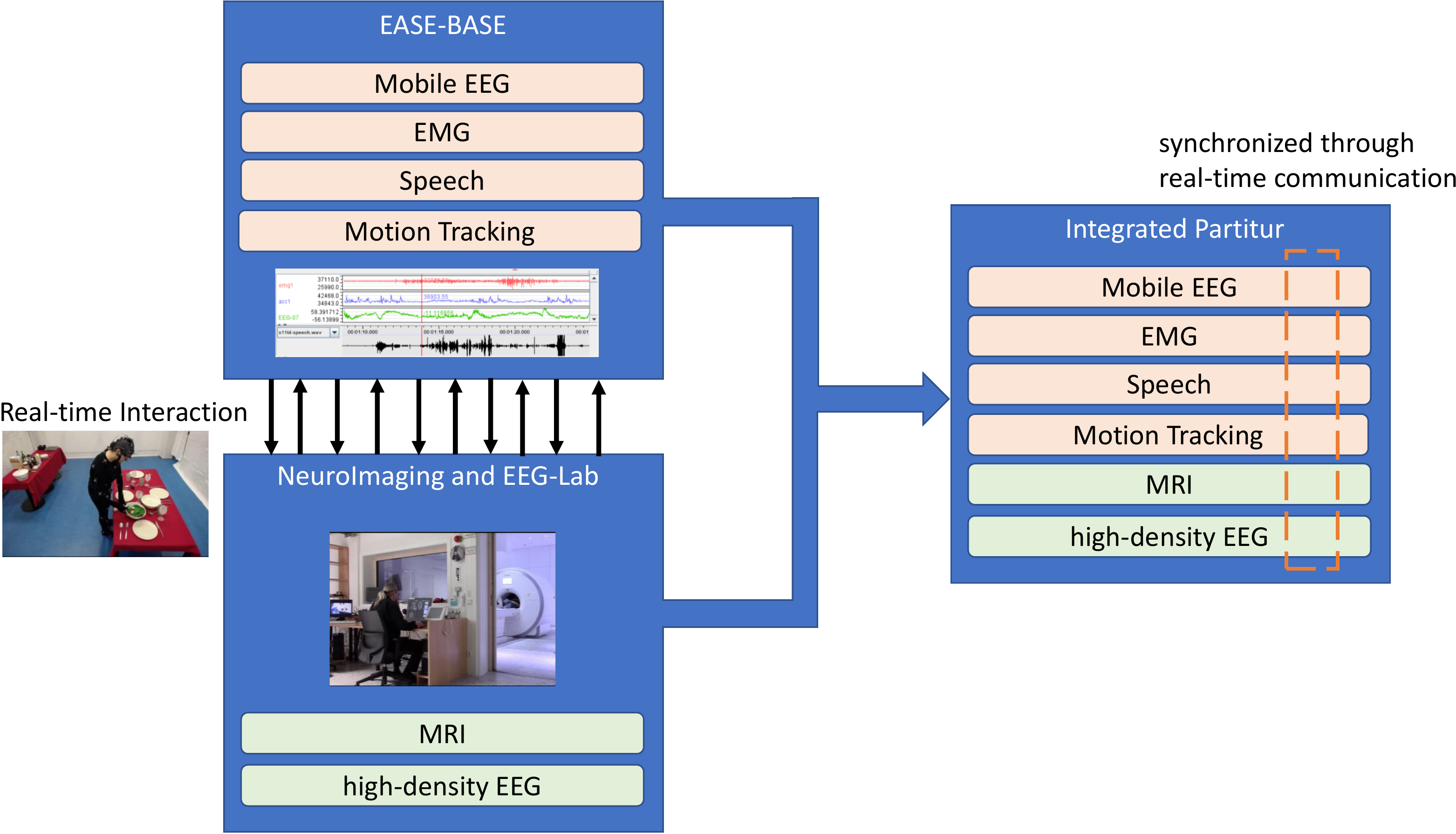}
\caption{LLL-3: Synchronous / Real-time LabLinking}
\label{fig:ll_online}
\end{figure}

In this section, we describe how we created a setup for performing low-latency aka real-time LabLinking between the Biosignals Lab and the Neuroimaging Lab, as shown in Figure~\ref{fig:ll_online}. For this purpose, we set up a video stream from one or multiple cameras in the Biosignals Lab to the Neuroimaging Lab where a second participant observed these video projected into the MRI scanner through a mirrored monitor screen inside the MRI tube. For time-synchronized recording and interpretation of signals, CSL develops extensive software suites, such as the recording software based on the middleware of ``Lab Streaming Layer (LSL)''. 

Real-time LabLinking required the setup of a number of technical prerequisites: First, we need to provide a low-latency network connection that requires real-time transmission of high resolution video (through a video streaming connection based on gstreamer (\url{https://gstreamer.freedesktop.org/})) between the participating laboratories. This also involved setting up the necessary permissions and tunnels through the universities network architecture. Second, the image transmitted from the Biosignals Lab to the fMRI needed to be modified to fulfill not only the requirements but also to function as a sensor in the Biosignals Lab and to function as an appropriate stimulus to the participant in the MRI scanner. For this purpose, we explored two different options: A first-person perspective recorded from the world camera of the first participant's head-mounted eye tracker (pupil labs) and an overhead view of the whole scene from a camera mounted at the ceiling of the lab. Third, we needed to implement multi-channel acoustic feedback channels for real-time communication between all involved parties,  the participants and scientists who are conducting the experiments.

To demonstrate a real-time interaction paradigm through LLL-3, we created a food choice variant of the table setting scenario. Here, we modify the table setting scenario to include a number of decisions between more and less healthy food choices for setting up a breakfast table (e.g., white-flour toast vs.\ whole-grain bread). The participant in the Biosignals Lab is physically manipulating and inspecting the potential food items. The participant in the fMRI can perceive the same information through the video from head-mounted camera. By using response buttons inside the fMRI scanner, they can backchannel decisions on which food to take to the acting participant. Note that the roles of the two participants in this setup are different than for the original one: The participant in the MRI scanner is no longer asked to mirror the activities of the active person, but can actively influence the course of the experiment. This scenario shows that LLL-3 supports different role distributions between participants, depending on the desired measurement. In an expanded version of the present scenario these food choices are also triggered by color labeled food items, e.g., using a nutri score. 

\subsection{A Case of LLL-4 - Augmented and Virtual Realities}

The next paramount and timely extension of LabLinking concerns the establishment of a maximal immersive experience to ameliorate the spatial distance between participants. A high degree of immersion is crucial to ensure that the behavior and involved cognitive processes are as comparable as possible between a conventional and a LabLinking experiment. Recently, computer-generated interactive environments have been established in research and development in which users can view and perceive their reality together with physical properties - so-called virtual realities (VR). Virtual realities can be mixed with physical reality, which is then referred to as mixed or augmented (AR). A/VR technologies make it possible to embed users in different worlds (immersion) and let them carry out actions that have an impact on the virtual world (interactivity). This may give A/VR users the illusion that their interactions have an impact on the real world, i.e., that what seems to happen in the virtual world actually happens~\cite{slater2016enhancing}. A/VR thus provide fundamental tools and mechanisms for LLL-4 to enable users to interact with each other in common worlds while operating and manipulating machines and their environment. 

Both AR and VR require tracking of the participants to register their location and movements. Tracking is important to simulate the person's motion in the view of the other person or to position virtual elements relative to the participant. Just as for the cameras in the proposed design, we can re-use the sensors of the data collection for this purpose. For example, we can employ marker-based (OptiTrack) tracking or markerless tracking (with depth cameras). Both types of tracking systems are already in place in the Biosignals Lab setup.

\begin{figure}[!htbp]
  \centering
  \begin{minipage}[b]{0.45\columnwidth}
    \includegraphics[width=\textwidth]{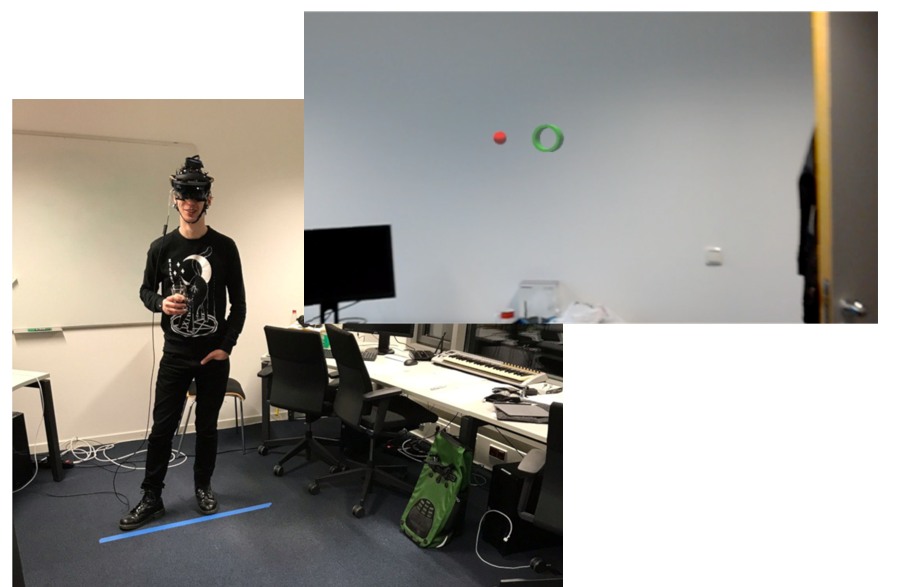}
    \begin{center}
        (a) AR setup with multimodal observation of attention~\cite{vortmann_eeg-based_2019}.
    \end{center}
  \end{minipage}
  \hfill
  \begin{minipage}[b]{0.45\columnwidth}
    \includegraphics[width=\textwidth]{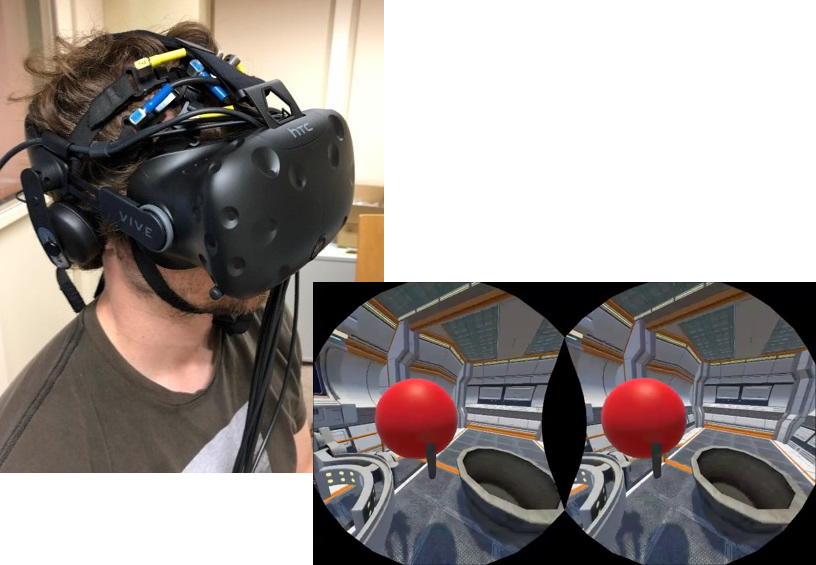}
    \begin{center}
        (b) VR setup with multimodal observation of mental workload~\cite{putze_decoding_2019}.
    \end{center}
  \end{minipage}
    \caption[Exemplary images captured by the two RealSense cameras]{LLL-4: Two examples of using AR/VR technology in combination with multimodal observation of participants.}
    \label{fig:arvr}%
\end{figure}

Despite the great possibilities, A/VR has an important shortcoming, i.e.\ it lacks the social context of local interaction. In particular, the virtual world only sees the projection of users and their immediate actions but not those social events that take place in parallel, such as incoming phone calls or conversations with bystanders. Technical interaction systems therefore require a robust assessment of the users' behavior. Systems need to decide if the observed behavior (or its correlates) relates to the interaction represented in A/VR. The lack of social context in A/VR technologies is not only a shortcoming but also an opportunity: A/VR enable to decouple individual components of the interaction, i.e.\ social signals could be systematically varied in virtual worlds. This paves the road to validate and predict the impact of social signals on interaction in a more rigorous, systematic, and ecological way as was possible in the past. An important tool to establish this research is the multimodal observation of participants and the processing of the recorded biosignal data through machine learning models. Previous work has shown the feasibility of real-time classification of workload, attention, or affect and the available sensors in the LabLinking setup already provide the necessary data. Augmenting the real or virtual experiment setup with such information in LLL-4 can support taking the other person's perspective or to interact with them naturally~\cite{oh2018systematic}. Figure~\ref{fig:arvr} shows two examples of using AR and VR together with multimodal observation of participants which will be integrated in planned LLL-4 setups.

\section{Conclusion \& Outlook}

LabLinking is not limited to the constellation described in this paper, but rather we can think of various other line-ups: For example, LabLinking can be employed to overcome bottlenecks in the form of available sensors. Well-equipped laboratories like the KD2Lab (\url{https://www.kd2lab.kit.edu}) which features 40 cabins each fitted with various biosignal sensors for large-scale interaction experiments may benefit from the option to include specialized devices  and experience, such as EEG or stationary equipment like fMRI. Sharing of research resources through LabLinking allows a more sustainable use of expensive sensor equipment and expert time: Individual labs can concentrate on their core strengths and link in needed additional modalities. Furthermore, LabLinking facilitates inter-cultural research as it makes it easier to connect labs in different countries or on different continents. Another way in which LabLinking can benefit the research community is that it encourages scientists to provide and support open and universal formats for data storage and data transmission, and it enables to expose students to latest technologies and equipment, and to integrate them into collaborative studies. These open source and sharing strategies are contributions towards effective open science practices.

\section*{Acknowledgments}
The research reported in this paper has been supported by the German Research Foundation DFG, as part of Collaborative Research Center (Sonderforschungsbereich) 1320 “EASE - Everyday Activity Science and Engineering”, University of Bremen (http://www.ease-crc.org/). The research was conducted in subproject H03 ``Descriptive models of human everyday activity''.


%
%
%

\bibliography{ieeeconf-LabCombi}

\begin{thebibliography}{10}

\bibitem{schultz2017biosignal}
Schultz T, Wand M, Hueber T, Krusienski DJ, Herff C, Brumberg JS.
\newblock Biosignal-based Spoken Communication: A Survey.
\newblock IEEE/ACM Transactions on Audio, Speech and Language Processing.
  2017;25(12):2257--2271.
\newblock doi:{10.1109/TASLP.2017.2752365}.

\bibitem{haidu_automated_2019}
Haidu A, Beetz M.
\newblock Automated models of human everyday activity based on game and virtual
  reality technology.
\newblock In: 2019 {International} {Conference} on {Robotics} and {Automation}
  ({ICRA}). IEEE; 2019. p. 2606--2612.

\bibitem{meier2018}
Meier M, Mason C, Porzel R, Putze F, Schultz T.
\newblock Synchronized Multimodal Recording of a Table Setting Dataset.
\newblock In: IROS 2018: Workshop on Latest Advances in Big Activity Data
  Sources for Robotics \& New Challenges, Madrid, Spain; 2018.

\bibitem{marsh_is_2013}
Marsh WE, Hantel T, Zetzsche C, Schill K.
\newblock Is the user trained? {Assessing} performance and cognitive resource
  demands in the {Virtusphere}.
\newblock In: 2013 {IEEE} {Symposium} on {3D} {User} {Interfaces} ({3DUI});
  2013. p. 15--22.

\bibitem{fehr_neural_2018}
Fehr T, Staniloiu A, Markowitsch HJ, Erhard P, Herrmann M.
\newblock Neural correlates of free recall of “famous events” in a
  “hypermnestic” individual as compared to an age- and education-matched
  reference group.
\newblock BMC Neuroscience. 2018;19(1):35.
\newblock doi:{10.1186/s12868-018-0435-y}.

\bibitem{peukert_acceptance_2019}
Peukert C, Pfeiffer J, Meissner M, Pfeiffer T, Weinhardt C.
\newblock Acceptance of {Imagined} {Versus} {Experienced} {Virtual} {Reality}
  {Shopping} {Environments}: {Insights} from {Two} {Experiments}.
\newblock In: 27th {European} {Conference} on {Information} {SystemsEuropean}
  {Conference} on {Information} {Systems}. ScholarSpace/AIS Electronic Library
  (AISeL); 2019. p. 1--16.

\bibitem{mason_iros2020}
Mason C, Gadzicki K, Meier M, Ahrens F, Kluss T, Maldonado J, Putze F, Fehr T,
  Zetzsche C, Herrmann M, Schill K, Schultz T.
\newblock From Human to Robot Everyday Activity.
\newblock In: IROS 2020. Las Vegas, USA: IEEE; 2020.Available from:
  \url{https://www.csl.uni-bremen.de/cms/images/documents/publications/mason_iros2020.pdf}.

\bibitem{gloy2020decision}
Gloy K, Herrmann M, Fehr T.
\newblock Decision making under uncertainty in a quasi realistic binary
  decision task--An fMRI study.
\newblock Brain and Cognition. 2020;140:105549.

\bibitem{Trautmann-Lengsfeld2013}
Trautmann-Lengsfeld SA, Dom{\'{\i}}nguez-Borr{\`{a}}s J, Escera C, Herrmann M,
  Fehr T.
\newblock The Perception of Dynamic and Static Facial Expressions of Happiness
  and Disgust Investigated by {ERPs} and {fMRI} Constrained Source Analysis.
\newblock {PLoS} {ONE}. 2013;8(6):e66997.
\newblock doi:{10.1371/journal.pone.0066997}.

\bibitem{wittenburg2006}
Wittenburg P, Brugman H, Russel A, Klassmann A, Sloetjes H.
\newblock ELAN: a Professional Framework for Multimodality Research.
\newblock In: Proceedings of the Fifth International Conference on Language
  Resources and Evaluation (LREC’06); 2020.

\bibitem{Jeannerod1995}
Jeannerod M.
\newblock Mental imagery in the motor context.
\newblock Neuropsychologia. 1995;33(11):1419--1432.
\newblock doi:{10.1016/0028-3932(95)00073-c}.

\bibitem{slater2016enhancing}
Slater M, Sanchez-Vives MV.
\newblock Enhancing our lives with immersive virtual reality.
\newblock Frontiers in Robotics and AI. 2016;3:74.

\bibitem{oh2018systematic}
Oh CS, Bailenson JN, Welch GF.
\newblock A systematic review of social presence: Definition, antecedents, and
  implications.
\newblock Frontiers in Robotics and AI. 2018;5:114.

\bibitem{vortmann_eeg-based_2019}
Vortmann LM, Kroll F, Putze F.
\newblock {EEG}-{Based} {Classification} of {Internally}- and
  {Externally}-{Directed} {Attention} in an {Augmented} {Reality} {Paradigm}.
\newblock Frontiers in Human Neuroscience. 2019;13.
\newblock doi:{10.3389/fnhum.2019.00348}.

\bibitem{putze_decoding_2019}
Putze F, Herff C, Tremmel C, Schultz T, Krusienski DJ.
\newblock Decoding {Mental} {Workload} in {Virtual} {Environments}: {A} {fNIRS}
  {Study} using an {Immersive} n-back {Task}.
\newblock In: 2019 41st {Annual} {International} {Conference} of the {IEEE}
  {Engineering} in {Medicine} and {Biology} {Society} ({EMBC}); 2019. p.
  3103--3106.

\end{thebibliography}
\bibliographystyle{plos2015}

\end{document}